\documentclass[prl,floatfix,superscriptaddress,twocolumn,aps,showpacs,amsmath,amssymb]{revtex4}

\usepackage[latin1]{inputenc}

\usepackage{graphicx}
\usepackage{psfrag}
\graphicspath{{./}}
\newcommand{\ColorOnline}{(Color online) }

\newcommand{\ket}[1]{\ensuremath{\vert{#1}\rangle}}
\newcommand{\bra}[1]{\ensuremath{\langle{#1}\vert}}

\newcommand{\eh}{\frac{1}{2}}

\newcommand{\NDot}{\ensuremath{N_{\mathrm{Dot}}}}
\newcommand{\Vg}{\ensuremath{V_{\mathrm{Gate}}}}
\newcommand{\VgOp}{\ensuremath{\hat{V}_{\mathrm{Gate}}}}

\newcommand{\HDot}{{\ensuremath{\cal H}_{\mathrm{Dot}}}}

\newcommand{\TrafoU}{{\ensuremath{\cal U}}}

\input rotate

\begin{document}

\title{The dark side of DFT based transport calculations}

\author{Peter Schmitteckert}%
  \affiliation{Institute of Nanotechnology, 
    Karlsruhe Institute of Technology, 76344 Eggenstein-Leopoldshafen,
    Germany}
\affiliation{DFG Center for Functional Nanostructures, Karlsruhe Institute of Technology, 76128 Karlsruhe, Germany}
\date{\today}
\begin{abstract}
We compare the conductance of an interacting ring of six lattice sites
threaded by flux $\pi$ in a two terminal setup with the conductance 
of the corresponding Kohn-Sham particles. 
Based on symmetry considerations we can show that even within (lattice) 
Density Functional Theory employing the exact Functional the conductance of the
Kohn-Sham particles is exactly zero, while the conductance of the physical system 
is close to the unitary limit. We show that this fundamental problem might be
solved by extending the standard DFT scheme.
\end{abstract}
\pacs{31.15.Ar, 71.15.Mb, 81.07.Nb}
\maketitle
Running an electrical current through individual molecules and 
being able to control the current flow by molecular design 
is one of the intriguing aspects of Molecular Electronics. 
In order to model the transport properties of a molecule
one has to take the contact region into account. Leading to
the problem of many degrees of freedom.
The ``standard method'' is a combination of Kohn-Sham (KS) density 
functional theory (DFT) calculations and the (self consistent) 
Landauer approach. \cite{Brandbyge:PRB02,Evers_Weigen_Koentopp:PRB2004,Arnold_Weigend_Evers:JCP2007}
At least for larger molecules, comprising  
a few hundred electrons, the standard approach appears to be 
without computationally feasible alternative, at present. 
Recently, the question to which extend this approach is actually justified
got increased attention.
At least for the the linear conductance, which can be obtained from
a ground state correlation function \cite{Bohr_Schmitteckert_Woelfle:EPL2006,Bohr_Schmitteckert:PRB2007},
one can hope that there exists a one to one mapping between the wave function and the 
conductance, from which a functional for the conductance would result by virtue of the 
Hohenberg-Kohn theorem \cite{Hohenberg_Kohn:PRB1964}. 
However, the nature of such a functional is not known and there is no reason that
this functional should coincide with the conductance of the Kohn-Sham auxiliary particles.
\par
In \cite{Schmitteckert_Evers:PRL2008} we studied the conductance of a linear chain of five strongly interacting sites
via the density matrix renormalization group method \cite{White:PRL1992,White:PRB1993}, where the conductance
was obtained from the Kubo formula \cite{Bohr_Schmitteckert_Woelfle:EPL2006,Bohr_Schmitteckert:PRB2007}.
By reverse engineering the Kohn-Sham potentials \cite{Gunnarson_Schoenhammer:PRL1986,Schoenhammer_Gunnarson_Noack:PRB1995,Schmitteckert_Evers:PRL2008} leading to
the same local densities as in the DMRG we could compare the conductance of the Kohn-Sham system
with the conductance of the physical system and found an excellent agreement close to the resonances
with larger deviations in the conductance valleys.\cite{Schmitteckert_Evers:PRL2008} The agreement could be traced back to the
existence of an at least approximate  Friedel sum rule, where the conductance is basically
given by $G= G_0 \sin^2( \pi \NDot  )$, where $G_0$ is the on-resonance conductance. 
The role of the Friedel sum rule was recently also studied in \cite{Troester:2009,Stefanucci_Kurth:PRL2011,Bergfield_EtAl:PRL2012,Troester_Schmitteckert_Evers:PRB2012}
in the context of the single impurity Anderson model, where it holds exactly.\cite{Langreth:PR1966}
Therefore, it provides a conductance functional\cite{Evers_Schmitteckert:X2013}, 
and since it also holds in the non-interacting case,
the conductance of Kohn-Sham particles obtained via the Landauer approach agrees with the physical conductance.
Remarkably, the spectral function can be arbitrarily wrong, 
despite an exact agreement between the physical and the Kohn-Sham conductance.\cite{Troester_Schmitteckert_Evers:PRB2012}
However, one should keep in mind that this is only valid within exact DFT. 
It was shown in \cite{Schenk_Schwab_Dzierzawa_Eckern:PRB2011} that evaluating the conductance within approximate functionals 
for the the above mentioned five site system can lead to parametrically wrong results, see also \cite{Schenk_Dzierzawa_Schwab_Eckern:PRB2008}. 
\par
\begin{figure}[t]
\begin{center}
    \includegraphics[width=0.475\textwidth]{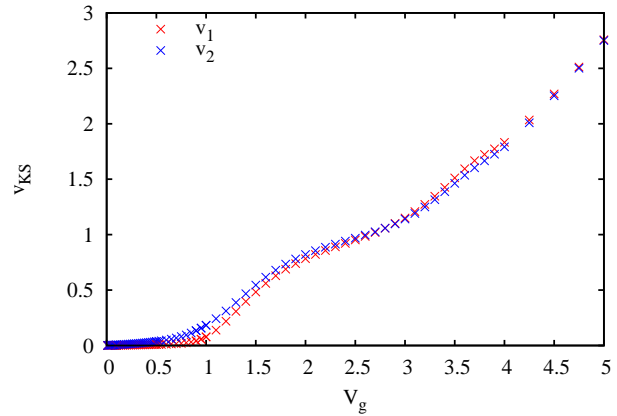}
    \caption{\ColorOnline Kohn-Sham potentials $v_1$ and $v_2$ obtained by reverse engineering
      the solution provided by DMRG calculations. The deviations, i.e. the asymmetric parts,
      are of the order of $10^{-7} J$ and below.}
   \label{fig:vKS}
\end{center}
\end{figure}

Here we compare the conductance of a hexagonal ring structure with the Kohn-Sham conductance 
obtained via exact DFT for a system, where the conductance is not given by a simple
Friedel sum rule. To this end we look at a six site ring of spinless fermions which
is threaded by a half magnetic flux quantum. 
The Hamiltonian of the structure is given by 
\begin{align}
 \HDot  &= - J \sum_{x=2}^{6}  \left( \hat{c}^+_{x} \hat{c}^{}_{x-1} + \hat{c}^{+}_{x-1} \hat{c}^{}_x \right) \nonumber
           + J \left(  \hat{c}^+_{1} \hat{c}^{}_{6} + \hat{c}^{+}_{6} \hat{c}^{}_1 \right)         \\
        &+   U  \sum_{x=2}^{6} \left( \hat{n}_{x} - \eh \right) \left( \hat{n}_{x-1} -\eh\right)   \label{eq:H} \\
        &+ U \left( \hat{n}_{1} - \eh\right) \left( \hat{n}^{}_{6} - \eh\right)                    \nonumber
\end{align}
where $J$ is the nearest neighbour hopping of the ring, $U$ is the nearest neighbour interaction
and the flux $\pi$ is modeled by changing the sign of the hopping element of the 1--6 bond.
The system is coupled via an hybridization of $J'$ from site 1 to a left and from site 4 to a right lead.
We are using exactly the same lead structure as in  \cite{Bohr_Schmitteckert:Ann2012} where each
lead starts with three real space sites and is then coupled to 34 sites represented in momentum space employing
a logarithmic discretization plus a linear discretization scheme close to the Fermi surface,
leading to a level spacing of the leads at zero energy of $7.6\cdot 10^{-5} J$, 
for details see \cite{Bohr_Schmitteckert:PRB2007,Bohr_Schmitteckert:Ann2012}.
The leads correspond to tight binding chains with hopping elements of $J$.
In the DMRG we kept 1000 states per block. In order to avoid getting stuck into excited states
we first performed an initial run for $U=0$ and $\Vg=0$. We then restarted the DMRG increasing first
the interaction $U$ in five steps to $U=2.25J$. Then we increased $\Vg$ performing 9 finite lattice sweeps
for each gate voltage shown in Figure~\ref{fig:vKS}.
\par 
This system has the interesting property that in the noninteracting limit its linear conductance 
is precisely zero for all applied gate voltages $\Vg$ \cite{Bohr_Schmitteckert:Ann2012},
\begin{equation}
 \VgOp= \Vg  \sum_{x=1}^{6} \hat{n}_x \,.
\end{equation}
In contrast, a nearest neighbour interaction of the order of $U=2t$ results in a geometric
Kondo effect leading to a conductance up to approximately 0.8 $e^2/h$, 
close to the unitary limit \cite{Bohr_Schmitteckert:Ann2012}.
Due to the particle hole symmetric form of the interaction $U$ in Eq.~\eqref{eq:H} 
each site is exactly half filled for $\Vg=0$ for any value of $U$. Since the conductance changes 
with respect to the interaction, while the densities remain constant,
 the conductance is not given by a Friedel sum rule.
Indeed, there is no reason why such a formula should exist, since the Friedel sum rule is a statement about the spectral
function, which in general has no simple relation to the conductance except for proportional coupling,
e.g.\ isolated levels.\cite{Meir_Wingreen:PRL1992}
\begin{figure}[t]
\begin{center}
    \includegraphics[width=0.45\textwidth]{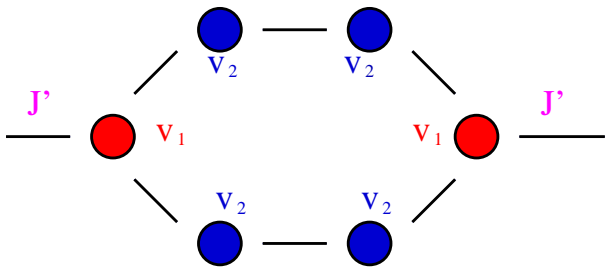}
    \caption{\ColorOnline  The most general set of parameter allowed by the symmetry of the system.
            $v_1$ and $v_2$ are the Kohn-Sham potentials of the system including the external gate voltage $\Vg$.}
    \label{fig:DFT}
\end{center}
\end{figure}

In Figure~\ref{fig:DFT} we display the most general structure of Kohn-Sham potentials
that is consistent with the symmetry of the structure. That is, the potentials of the contact
sites have to be identical due to inversion symmetry, and the middle four sites are subject to
the same potential due to inversion and mirror symmetries of the ring structure, leading to two
Kohn-Sham potentials $v_{1,2}$ only. In Figure~\ref{fig:vKS} we show the reverse engineered Kohn-Sham potentials
for a sample system with $J'=0.5J$. In the low gate voltage we see a compensation of the gate
voltage corresponding to the conductance plateau in the Kondo regime. For gate voltages of
the order of $U/2$ we see a cross over to a Hartree regime, which gets itself interrupted by
the appearance of the next pair of degenerate states. In the numerics we calculated the Kohn-Sham potentials
on all real space sites, and the deviation from the symmetry as displayed in the  Figure~\ref{fig:DFT} are
of the order of $10^{-7}$, which is in agreement with the numerical accuracy. 
By the explicit calculation of the Kohn-Sham potentials we can show that out system is indeed
$v$-representable, which means that a corresponding Kohn-Sham system does indeed exist.
In addition we do get Kohn-Sham potentials
on the lead sites. However, they only lead to a slight change of the lead Greens function.
\par
However, it is straightforward to show, that even for arbitrary potentials $v_1$ and $v_2$ the transmission
amplitude is still zero. Therefore, no matter what the actual values of the Kohn-Sham potentials are,
the conductance of the Kohn-Sham auxiliary system is always zero. And since it is always zero, it doesn't matter
that the lead Greens functions is changed by Kohn-Sham potentials induced in the noninteracting leads, as there 
are no conductance peaks which could be shifted or modified.
Finally, even the introduction of an an exchange correlation voltage,
as a consequence of non-adiabatic dynamic correlations stemming from the zero frequency limit of time-dependent DFT, \cite{PS_Dzierzawa_Schwab:PCCP2013} 
will not lead to a finite conductance, at least as long as the effective potentials are finite.

\begin{figure}[ht]
\begin{center}
    \includegraphics[width=0.45\textwidth]{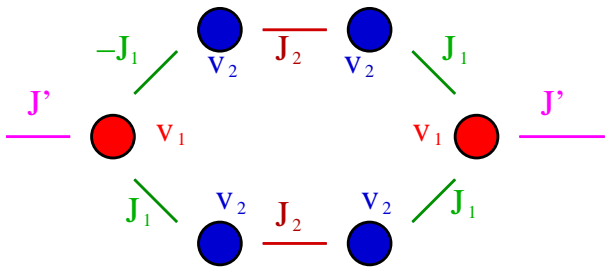}
    \caption{\ColorOnline  The most general set of parameter allowed by the symmetry of an extended DFT,
            where one also adapts the kinetic part of the Hamiltonian.}
    \label{fig:DFTKin}
\end{center}
\end{figure}

One may think that the above result is due to some extreme fine tuning of the model.
We therefore introduce Kohn-Sham hopping elements in such a way, that the kinetic energy of
bonds of the ring in the physical system and the auxiliary free fermion system are the same.
Again, the symmetry of the ring restricts the possible hopping elements to $J_1$ and $J_2$ as
displayed in Figure~\ref{fig:DFTKin}, and we can even allow for the adaption of the contact
term $J'$. Remarkably, even with this general set of parameter, the conductance of the auxiliary system
is exactly zero for all gate voltages.

In the preceding part we have shown an example where the conductance of the Kohn-Sham particles 
fails to describe the real physical system. In the following we want to show that the single
particle description can be improved by considering a reduced single particle density matrix description.
To this end we perform a ground state DMRG calculation where we extract the single particle
reduced density matrix 
\begin{equation}
			K_{x,y} = \bra{\Psi_0} \hat{c}^{+}_x \hat{c}^{}_y \ket{\Psi_0} \label{eq:def_K} \,,
\end{equation}
where we restrict the site indexes to the structure and possibly including up to the first three sites
of the leads closest to the structure. 
The results presented below are obtained from DMRG calulations where we restricted each lead to 22 levels
in the momentum space description, in contrast to the 34 levels used in the reference calculation \cite{Bohr_Schmitteckert:Ann2012}.
By diagonalizing $K$,
\begin{equation}
  \tilde{K}_{\ell, \ell'} = f_\ell \, \delta_{\ell,\ell'} = \TrafoU^+ \cdot K \cdot \TrafoU \,
\end{equation}
we obtain the occupation number $f_\ell$ of the so called natural orbitals given by the transformation matrix $\TrafoU$.
In a next step we make the assumption that we can treat the natural orbitals as independent levels.
Note, that due to the symmetry of our structure we still get occupation numbers in degenerate pairs, 
and we can still have interference effects.

\begin{figure}[th]
\begin{center}
    \includegraphics[width=0.45\textwidth]{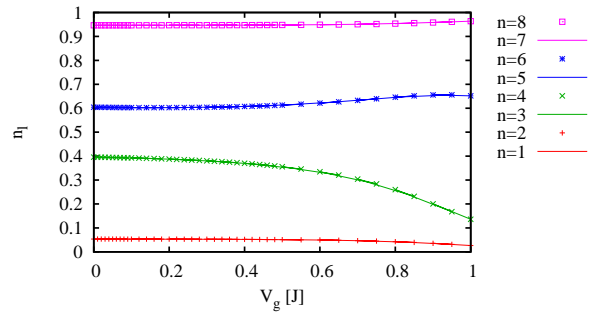}
    \caption{\ColorOnline The occupation $n_\ell$ of the natural orbital, where
               for the six site structure enhanced by the first lead site of each lead.
               The pluses (crosses, starts, boxes) correspond to level $\ell=2$ (4,6,8),
               where the lines correspond the  to the degenerate partner $\ell=1$ (3,5,7).}
    \label{fig:ROccupation}
\end{center}
\end{figure}
The occupation numbers in the case of restricting $K$ to the structure enhanced by the first lead site of each lead
are displayed in Figure~\ref{fig:ROccupation}. In contrast to the total dot occupation which is close to an integer, $N_{\mathrm{Dot}}=3$,
we now have four levels which are close to a half integer occupation and which can contribute of transmission.
Interestingly, there is an increase of the occupation of the four higher occupied levels despite the fact that we are pushing 
particle out of the structure by applying the gate voltage $\Vg$.
It seems natural to blindly add the conductances of each level obtained by assuming a Friedel sum rule contribution for each level.
However, due to the one-dimensional character of our leads, the complete conductance must not exceed $e^2/h$.
We therefore weight the individual contribution with a factor $w_\ell$,
\begin{equation}
   G = \sum_\ell w_\ell \sin^2\left( f_\ell \pi \right) \,, \label{eq:RFG}
\end{equation} 
where $w_\ell$ is set to the average occupation of the left- and right-most site of the structure under consideration, 
$w_\ell = \left( \TrafoU^2_{1,\ell} + \TrafoU^2_{M_{\mathrm S},\ell} \right)/2$,
where $M_{\mathrm S}$ is number of sites of the structure plus the added real space lead sites.
We would like to stress that we are not claiming that Eq.~\eqref{eq:RFG} combined with our choice of $w_\ell$
has to be a justified procedure. It is only meant to provide a first impression and to ensure that $G$ is
within the limits of zero and one conductance quantum. Using $w_\ell = \TrafoU^2_{1,\ell}$ or $w_\ell = \TrafoU^2_{M_{\mathrm S},\ell}$ 
leads to basically the same result.

\begin{figure}[th]
\begin{center}
    \includegraphics[width=0.45\textwidth]{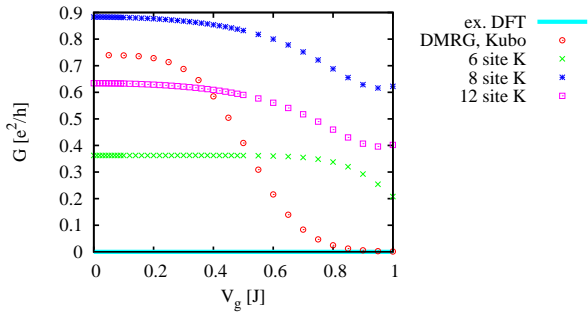}
    \caption{\ColorOnline Comparison of the conductance obtained via the Kubo formalism within DMRG (circles),
              and the one obtained from Eq.~\eqref{eq:RFG}, where the reduced density matrix $K$ was restricted
              to the six site structure (green crosses), the structure plus the 
              contact site of each lead (blue stars), and the structures plus the first 
               three sites of the leads on each site (magenta squares).}
    \label{fig:RFriedel}
\end{center}
\end{figure}

In Figure~\ref{fig:RFriedel} we compare the result of  Eq.~\eqref{eq:RFG} applied to the six site structures,
and the cases where we include one or three sites from each lead, to the Kubo result within DMRG \cite{Bohr_Schmitteckert:Ann2012}.
Remarkably, in contrast to the failure of the plain DFT approach our simple approach of Eq.~\eqref{eq:RFG} not only gives
a peak at all, it already provides the correct  order of magnitude for the peak height and the peak width. 
We would like to stress that within this description we are beyond a single slater determinant description.
Due to the non-integer occupation numbers we have actually moved to a multi reference description.

From this we conjecture that by going beyond a standard DFT description towards 
a reduced density matrix functional theory \cite{Gilbert:PRB1975,Pruschke:PRB2011}
or a generalized td-CDFT \cite{Tokatly:PRB2011}
one should be able to improve todays simulations within molecular electronics.
In summary we have provided an example where the conductance of the Kohn-Sham auxiliary particles
fails remarkably to describe the physical conductance. The Kohn-Sham conductance is 
strictly zero for all gate voltages, while the conductance of the physical system displays 
resonance peaks close to the unitary limit. In contrast, by considering a description based on
the eigenstates of a single particle reduced density matrix we can obtain results which are 
at least in the correct order of magnitude. Results presented in this work and the population
blocking mechanism \cite{Bohr_Schmitteckert:Ann2012} lead to the conclusion that one should
apply multi reference methods in order to describe transport properties interacting quantum system 
in the presence of interference effects
within a single particle description reliably.
Our finding suggest that a reduced single particle density matrix based approach is a promising
candidate to study transport within a mean field type method, i.e.\ a single particle, description.
\acknowledgments Useful discussions with Ferdinand Evers, Peter Schwab and Peter W{\"o}lfle are greatfully acknowledged. 
The Kubo based DMRG calculations have been performed at the XC2 
of the SSC Karlsruhe under the grant number RT-DMRG.
This work is supported by the German science foundations (DFG) within the priority programme SPP1423.

\bibliographystyle{apsrev}
\bibliography{DFT}

\end{document}